\documentclass[aps,pre,amsmath,amsfonts,floatfix,superscriptaddress,nofootinbib,twocolumn]{revtex4-1}
\pdfoutput=1
\usepackage{mathrsfs} \usepackage{amssymb} \usepackage{graphicx,color} \usepackage{bbm} \usepackage{multirow}
\usepackage{bbm} \usepackage{mathrsfs} \usepackage{commath}\usepackage{tikz}
\usetikzlibrary{shapes,arrows,positioning,calc}\usepackage{stmaryrd}\usepackage{bm}

\let\a=\alpha \let\b=\beta \let\g=\gamma \let\d=\delta
\let\e=\epsilon   
 \let\m=\mu   
\let\s=\sigma  \let\f=\varphi

   \let\io=\infty

 \def\HH{{\cal H}}
\def\NN{{\cal N}}

\def\to{\rightarrow}

\newcommand{\beq}{\begin{equation}} \newcommand{\eeq}{\end{equation}}

\begin{document}

\title{
Protocol-dependent shear modulus of amorphous solids
} 

\author{Daijyu Nakayama}
\affiliation{Cybermedia Center, Osaka University, Toyonaka, Osaka 560-0043, Japan}
\affiliation{Graduate School of Science, Osaka University, Toyonaka, Osaka 560-0043, Japan}

\author{Hajime Yoshino}
\affiliation{Cybermedia Center, Osaka University, Toyonaka, Osaka 560-0043, Japan}
\affiliation{Graduate School of Science, Osaka University, Toyonaka, Osaka 560-0043, Japan}

\author{Francesco Zamponi}
\affiliation{Laboratoire de Physique Th\'eorique,
ENS \& PSL University, UPMC \& Sorbonne Universit\'es, UMR 8549 CNRS, 75005 Paris, France}

\begin{abstract}
We investigate the linear elastic response of amorphous solids to a shear strain at zero temperature. 
We find that the response
is characterized by at least two distinct shear moduli. The first one, $\mu_{\rm ZFC}$, is associated with the linear response
of a single energy minimum. 
The second, $\mu_{\rm FC}$, is related to sampling, through plastic events, an ensemble of 
distinct energy minima. 
We provide examples of protocols that allow one to measure both shear moduli.
In agreement with a theoretical prediction based on the exact solution in infinite spatial dimensions, 
the ratio $\mu_{\rm FC}/\mu_{\rm ZFC}$ is found
to vanish proportionally to the square root of pressure at the jamming transition. 
Our results establish that amorphous solids are characterized by a rugged
energy landscape, which has a deep impact on their elastic response, as suggested
by the infinite-dimensional solution.
\end{abstract}

\maketitle

\section{Introduction}
Most solid state textbooks are almost entirely devoted to crystals~\cite{AM76}.
The reason is obvious: while the theory of crystals is fully developed,
the theory of amorphous solids (glasses, foams, granulars, etc.) 
is still very incomplete~\cite{BB11,Ca09}.
Crystals can be understood as perfect periodic lattices, around which 
particles perform small vibrations. This allows one to construct a low-temperature
harmonic expansion, and obtain all thermodynamic properties in terms of harmonic
excitations, i.e. phonons. Moreover, crystal flow (or plasticity) 
and melting
is mediated by defects
(mostly dislocations) that are also quite well understood~\cite{AM76}.

The situation is very different for glasses, which display all kind of anomalies
with respect to crystals: they show an enhanced low-frequency density of states (the so-called Boson Peak)~\cite{MS86},
leading to anomalous behavior of specific heat and thermal conductivity~\cite{Ph87}. Crucially for our study, they show irreversible
``plastic'' response to arbitrarily small perturbations~\cite{ML99,CR00,ML04,RTV11,HJPS15}:
during plastic events, some part of the system relaxes irreversibly to a new low-energy state
by crossing some low-energy barrier~\cite{Go69,HKEP11,BW09b,Wy12,MW15}.

These observations suggest the following picture: crystals can be thought as isolated minima of the potential energy,
around which a well-defined harmonic expansion can be performed, and that are separated from other minima by high
enough energy barriers~\cite{AM76}. On the contrary, glasses are ``fragile'' minima of the potential energy function: they are characterized
by many soft modes~\cite{WNW05}, the harmonic expansion thus works only at extremely low temperatures~\cite{SBOS11,IBB12,DLW15}, 
and very low-energy barriers separate
each glassy minimum from many other neighboring, and equivalent, glassy minima~\cite{Go69,He08,nature}. 
In this picture, it is natural that even a very small
perturbation destabilizes a glassy minimum and brings the system over a barrier to relax, 
irreversibly, to another minimum~\cite{Go69,Wy12,nature}.

\begin{figure}[b]
\includegraphics[width=0.9\columnwidth]{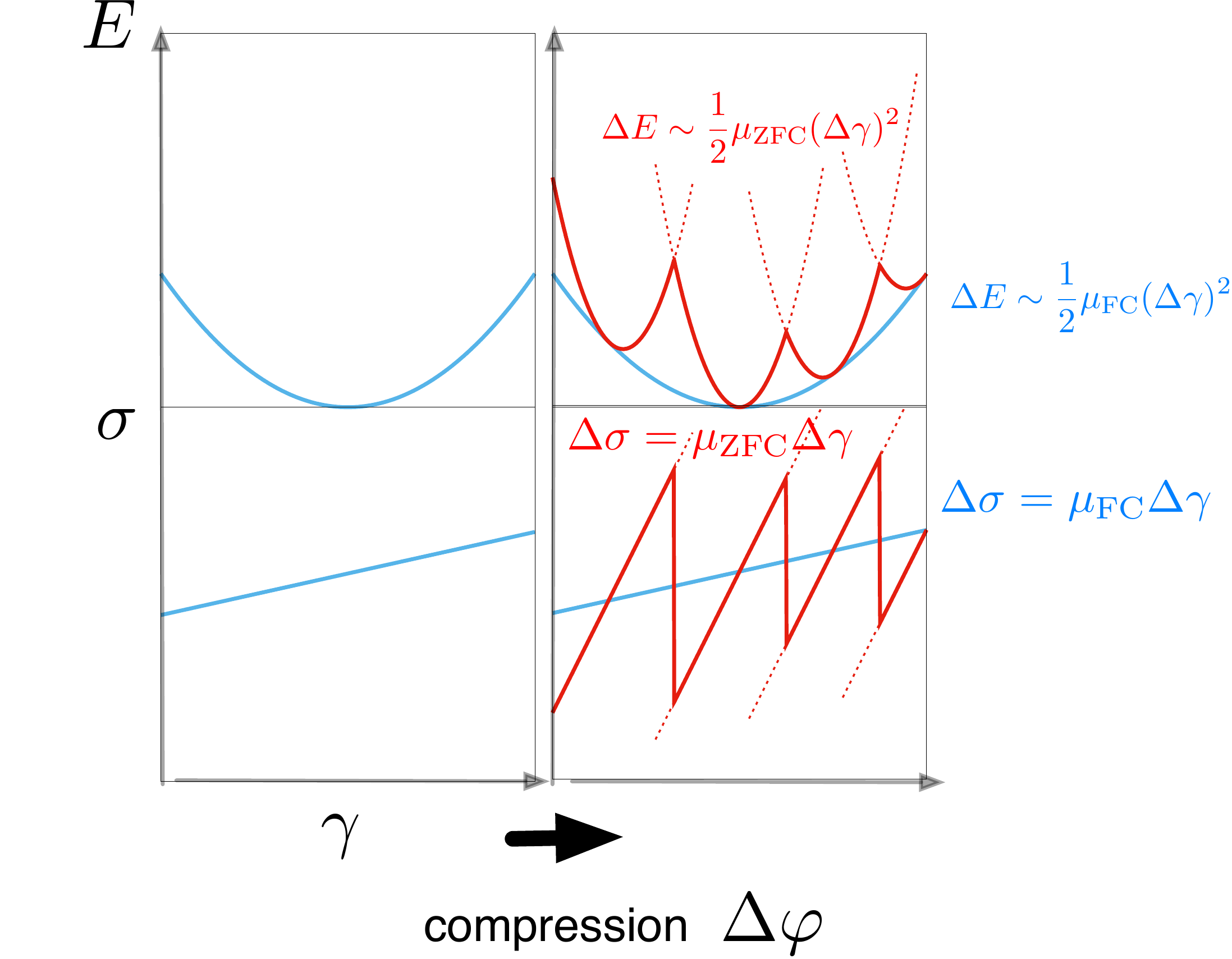}
\caption{
Oversimplified sketch of an energy landscape with two distinct shear moduli. 
{\it Top}: elastic energy $E$ versus shear strain $\g$.
{\it Bottom}: same illustration using stress $\s = dE/d\g$ as a function of strain.
Within a single energy minimum (left column), the energy increase behaves elastically
$\Delta E \propto (\Delta \gamma)^2$ for a small enough increment of the shear strain $\Delta \gamma$.
When the energy landscape becomes bifurcated (right column),
energy minima (red line) are organised in basins (blue line).
Each of the dotted line represents a region where a minimum is locally stable.
If the system can sample the basin, a lower shear modulus $\m_{\rm FC}$ is observed in the $\g\to0$ regime, corresponding to the envelope of the individual basins. The softening is due to inter-basin transitions.
}
\label{fig:land}
\end{figure}

The exact mathematical solution of the problem in the abstract limit of an infinite-dimensional space
can be the source of inspiration about some physical properties of the solid in three dimensions~\cite{PZ10,ARCMP16}.
In particular, it suggests
that the organisation of the energy minima is hierarchical~\cite{nature}: glassy minima are organised in clusters,
or ``basins'', themselves organised in larger basins, and so on, as it is well-known to happen in mean field 
spin glasses~\cite{MPV87,Pa06}. 
In such a situation, the response of the glass
to an external perturbation depends on how much of the energy landscape can be explored~\cite{MPV87,Pa06,Yo12,YZ14}. 
Consider elastic response. If only a given energy minimum
is explored, the system responds linearly with certain elastic coefficients. If a larger cluster of minima can be explored,
the response is still linear, but elastic coefficients are different (see Fig.~\ref{fig:land} for an illustration). 
Precise computations can be performed 
in the infinite-dimensional limit~\cite{YZ14,RU15}.

In this paper, inspired by this idea,
we explore the elastic response of
the simplest amorphous solid, a zero-temperature jammed assembly of soft spheres at pressure $P$,
to the simplest perturbation, shear strain. We focus on the vicinity of the jamming transition, which happens at
the density where
$P=0$ for the first time upon decompression~\cite{OSLN03}.
By analogy with spin glasses~\cite{MPV87}, we use two distinct measurement protocols to determine the linear
shear modulus:  in the ``zero-field compression'' (ZFC) protocol, one first reaches the target pressure and then applies 
the shear strain; in the ``field compression'' (FC) one first applies the shear strain,
and then compresses to target pressure $P$. The terminology comes from spin glasses where the
strain is replaced by a magnetic field~\cite{MPV87}.

 We obtain three main results.
{\it (i)} We show that in the ZFC protocol, the response is elastic with a shear modulus $\mu_{\rm ZFC}$ 
that characterizes a single glassy minimum. In the FC protocol the response turns out to be
still elastic  but with a distinct shear modulus $\mu_{\rm FC} < \mu_{\rm ZFC}$
due to plastic events or inter-valley transitions, similarly to what happens
with magnetic susceptibility in spin glasses~\cite{MPV87,BY86,nagata1979low,Pa06}.
This result suggests a non-trivial organisation of glassy
minima (but does not prove that it is hierarchical as in the infinite-dimensional solution).
{\it (ii)}~Infinite-dimensional calculations predict that in the limit in which the solid unjams, and $P\to 0$, the hierarchical organisation of basins
becomes fractal~\cite{nature}; in this limit, it is predicted that $\mu_{\rm ZFC} \propto P^{1/2}$ while $\mu_{\rm FC} \propto P$, thus 
$\mu_{\rm FC} \ll \mu_{\rm ZFC}$ resulting in a sharp separation of the two shear moduli~\cite{YZ14}. 
Our numerical data
agree with the theoretical prediction.
{\it (iii)}
We find that $\mu_{\rm FC}$ decreases with increasing system size, suggesting that $\mu_{\rm FC}=0$ in
the thermodynamic limit. This finding is not consistent with the most naive expectation based on the infinite-dimensional solution, and could
be due to several aspects of our numerical simulation protocol, as we discuss below.

\section{Methods}

\subsection{Details of the system}

We study a $3$-dimensional system of $N=1000-4000$
particles interacting via a soft repulsive contact pair potential
\beq
U = \sum_{i  < j} \phi_{ij}(r_{ij}) \ ,
\eeq
where $r_{ij}=|{\bf r}_{ij}| = |{\bf r}_{i}-{\bf r}_{j}|$ is the distance between particles $i$ and $j$,
and $\phi_{ij}(r)=\epsilon(1-r/D_{ij})^{2}$ for $r<D_{ij}$ and zero otherwise.
Here $D_{ij}=(D_{i}+D_{j})/2$
where $D_{i}$ is the diameter of the $i$-th particle.
To avoid crystallization, we consider a binary mixture of $N/2$ particles with diameter $D_{1}$ 
and $N/2$ particles with diameter $D_{2}$ with the ratio $D_{2}/D_{1}=1.4$. This is a standard choice in studies
of jamming~\cite{OSLN03,BW09}. 

In thermal equilibrium, the control parameters are reduced temperature $\hat T = k_B T/\e$ and
volume fraction $\varphi=(\pi/12)(D_1^{3} + D_2^{3})\rho$, where 
$\rho=N/V$ is the number density and $V$ is the volume of the system.
Note that 
inflating the particles by increasing the diameters $D_{1,2}$ 
is completely equivalent to reducing the volume $V$:
both operations amount to an increase of $\varphi$ at constant $\hat T$, i.e. a compression.
The main observables we consider are
pressure, which is the response of the system to a change in its volume:
\beq\label{eq:P}
P=-\frac1{3V}\sum_{i<j}{\bf r}_{ij} \cdot \nabla \phi_{ij}(r_{ij}) \ ,
\eeq
and shear-stress, which is the response to a volume-preserving change of boundary conditions
corresponding to a shear-strain:
\beq\label{eq:sigma}
\sigma=\frac1V \sum_{i<j} x_{ij} z_{ij} (\phi_{ij}'(r)/r)_{r=r_{ij}} \ ,
\eeq
where $x_{ij}$, $z_{ij}$
are $x$ and $z$ components of the vector ${\bf r}_{ij}$. 
Eqs.~\eqref{eq:P} and \eqref{eq:sigma} provide the microscopic expressions of pressure and shear-stress for
a given particle configuration; these expressions must be averaged over an appropriate ensemble of configurations,
as described below.
Throughout the paper,
$\e$ and $D_1$ are used as units of energy and length.

\subsection{Preparation of the samples}
\label{sec:preparation}

Each of our $\NN_s = O(10^{4})$ independent ``samples'' is obtained as follows. 
We start by a random configuration at $\varphi_{\rm init}=0.64$
and we run molecular dynamics (MD) simulation at $\hat T=10^{-5}$
for $30\tau_{\text col}$, where $\tau_{\text col}$ is the typical collision time.
This is done in order to stabilyze the system against small thermal fluctuations within the initial
energy basin selected by the random configuration, similarly to~\cite{IBB12}.
However, note that $\hat T=10^{-5}$ is such a low temperature that no barrier crossing to other
glassy basins can occur, so our system effectively remains trapped into a random energy basin,
selected by the initial random configuration~\cite{BW09,IBB12}. This will be a crucial observation
for the following discussion.

Next, we bring the system to $\hat T = 0$ by energy minimization 
using the conjugated gradient (CG) method~\cite{press2007numerical},
i.e. we reach an energy minimum close to the thermally stabilized configurations.
We obtain in this way our initial configurations at $\hat T=0$ and $\varphi_{\rm init}$,
and from now on we always work at zero temperature.
Note that $\f_{\rm init}=0.64$ is lower than the jamming density $\f_j\approx 0.6466$~\cite{OSLN03,IBB12} 
and thus the initial configurations are unjammed, i.e. they have
zero pressure.

\subsection{Measurement protocols}

To each sample we then apply two different protocols to measure the shear modulus, inspired by the ones
used in spin glasses~\cite{nagata1979low,BY86,Pa06}.
They consist in compressing the samples in presence or in absence of a shear strain.

Before describing the protocols, we specify that
(de)compression 
is done in small steps, during which the system is subjected to {\it (i)} affine deformation (multiplying by a common factor 
all particles' diameters in such a way that $\f$ changes by an amount 
$d\f = 5.0 \times 10^{-3}$) followed by {\it (ii)}
energy minimization via CG. 
Shear strain $\gamma$ is also applied in two steps by
{\it (i)} affine deformation, where
$x_{i} \to x_{i}+\gamma z_{i}$ for all particles (boundary condition into the $z$ direction
are also shifted by the Lees-Edwards scheme~\cite{LE72}), followed by {\it (ii)} energy minimization via~CG.

In the Field Compression (FC) protocol, 
the system is first subjected to a shear $\gamma$ at $\f_{\rm init}$.  
Then it is adiabatically compressed (AC) in small steps 
(affine deformation + CG)
up to $\f_{\rm f}=0.66$ corresponding to a pressure $P_{\rm f} \simeq 0.014$.
The remanent shear stress $\sigma(P,\g)$
is measured at fixed values of the pressure $P\in [0,P_{\rm f}]$, 
and from it we deduce the FC shear modulus $\mu_{\rm FC}(P,\g)=\sigma(P,\g)/\gamma$.
Next, the system is adiabatically
decompressed (AD) back to $\f_{\rm init}$ and the same measurements are performed
along the way.
In the Zero-Field Compression (ZFC), 
the system is AC up to the same $P_{\rm f}$ 
and then AD in small steps in absence of any shear.
The stress and pressure are measured after each step of the compression
and decompression. To measure the stress, in the ZFC case we take the current configuration 
and apply to it a small strain $\g$, and measure $\mu_{\rm ZFC}(P,\g)=\sigma(P,\g)/\gamma$;
the sheared configuration is then discarded.

In both cases, the averages over different samples are done at constant
pressure and not at constant $\f$: 
in fact, due to finite-size effects, the jamming point $\f_j$ where
pressure vanishes depends on the sample~\cite{OSLN03}. If we want to study
the scaling for $P\to 0$ it is better to average at constant pressure than at constant density.
In practice, averaging over the samples with a given pressure
$P$ is done by collecting data in the range $[P,P+dP]$ choosing some $dP$.
We examined $dP$ in the range $O(10^{-5})-O(10^{-3})$
and found that its precise choice is irrelevant: 
here we choose it such that we have a good number of samples in each pressure bin.

In the ZFC process we also measure the shear modulus directly at $\g=0$
via the ``fluctuation formula''~\cite{ML04}:
\begin{equation}\label{eq:ffZFC}
\mu_{\rm ZFC}(P,\g=0)=b -  \frac{1}{V}\sum_{i=1}^{N} {\bf \Xi}_{i} \cdot ({\cal H}^{-1} {\bf \Xi})_{i} \ .
\end{equation}
Here $b$ is the Born term (affine part of $\m$) defined as
\beq
 b=\frac{1}{V}\sum_{i < j} \left(z_{ij}\frac{\partial}{\partial x_{ij}}\right)^{2}\phi(r_{ij})
\eeq
while the second term is the non-affine correction, defined by
the Hessian matrix
\beq\nonumber
\HH_{ij}^{\mu \nu}
=\delta_{\mu \nu}
\delta_{ij}\sum_{k=1}^{N} 
\phi^{\mu \mu}(r_{ik})
-\phi^{\mu \nu}(r_{ij}) \ , \  \
\phi^{\mu \nu}(r_{ij})\equiv  \frac{\partial^{2}\phi(r_{ij})}{\partial x^{\mu}_{i}\partial x^{\nu}_{j}}
\eeq
where $\mu,\nu=x,y,z$, and
${\bf \Xi}_{i}= \nabla_{i} \sigma $.
Note that this zero-temperature formula assumes purely harmonic response 
excluding any plasticity. Its finite-temperature version, on the contrary,
can take into account all kinds of thermal excitations including plastic ones~\cite{Yo12}.
Note also that, unfortunately, there is no analog of the fluctuation formula for the FC measurement:
in fact, while the linear ZFC shear modulus is a property of a single configuration and can thus be written
as in Eq.~\eqref{eq:ffZFC}, in the FC case the linear shear modulus is a property of the whole basin and
for that reason a fluctuation formula necessarily involves an averaging over different minima in a basin
with weights that are difficult to determine~\cite{YZ14}.

\begin{figure}[t]
\includegraphics[width=0.8\columnwidth]{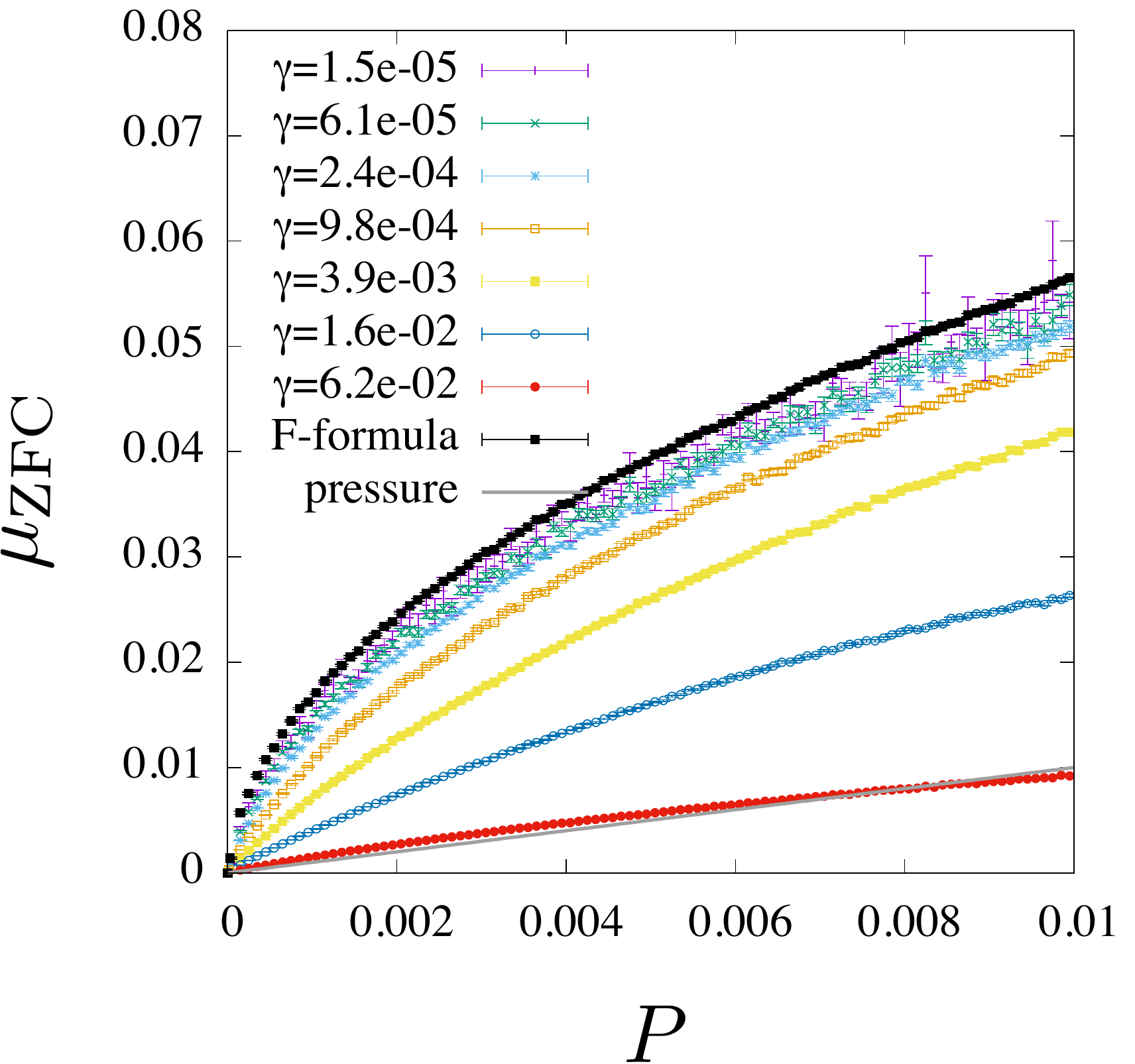}
  \caption{ZFC shear modulus, for which AC and AD give indistinguishable results.
$N=1000$ is used for the analysis of $\mu_{\rm ZFC}$.
Data with various values of the strain $\g=2^{-9},2^{-8},\ldots,2^{-3}$ are shown.
$dP=10^{-4}$ is used for binning.
  	The number of samples for each bin is $O(10^{4})$.
 ``F-formula'' indicates data obtained via fluctuation formula.
 }
\label{fig:ZFC}
\end{figure}

\begin{figure}[t]
\includegraphics[width=0.8\columnwidth]{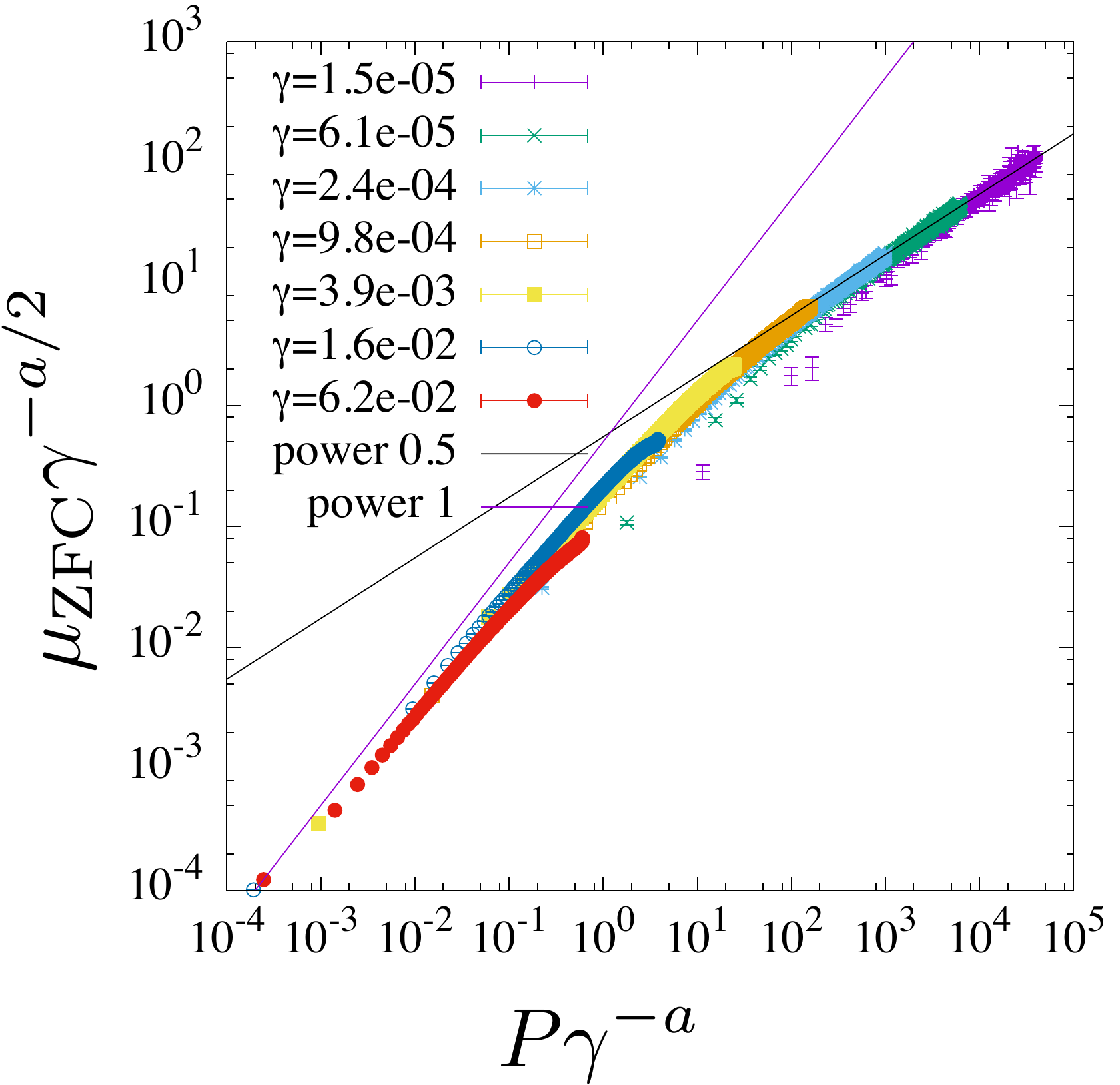}
\caption{ZFC shear modulus, scaled according to Eq.~\eqref{eq:ZFCsc} with $a=4/3$.
  	Here $dP=2.5\times10^{-5}$ and 
	$O(10^{3})$ samples are in each bin. 
}
\label{fig:ZFCscaled}
\end{figure}

\begin{figure}[t]
\includegraphics[width=0.8\columnwidth]{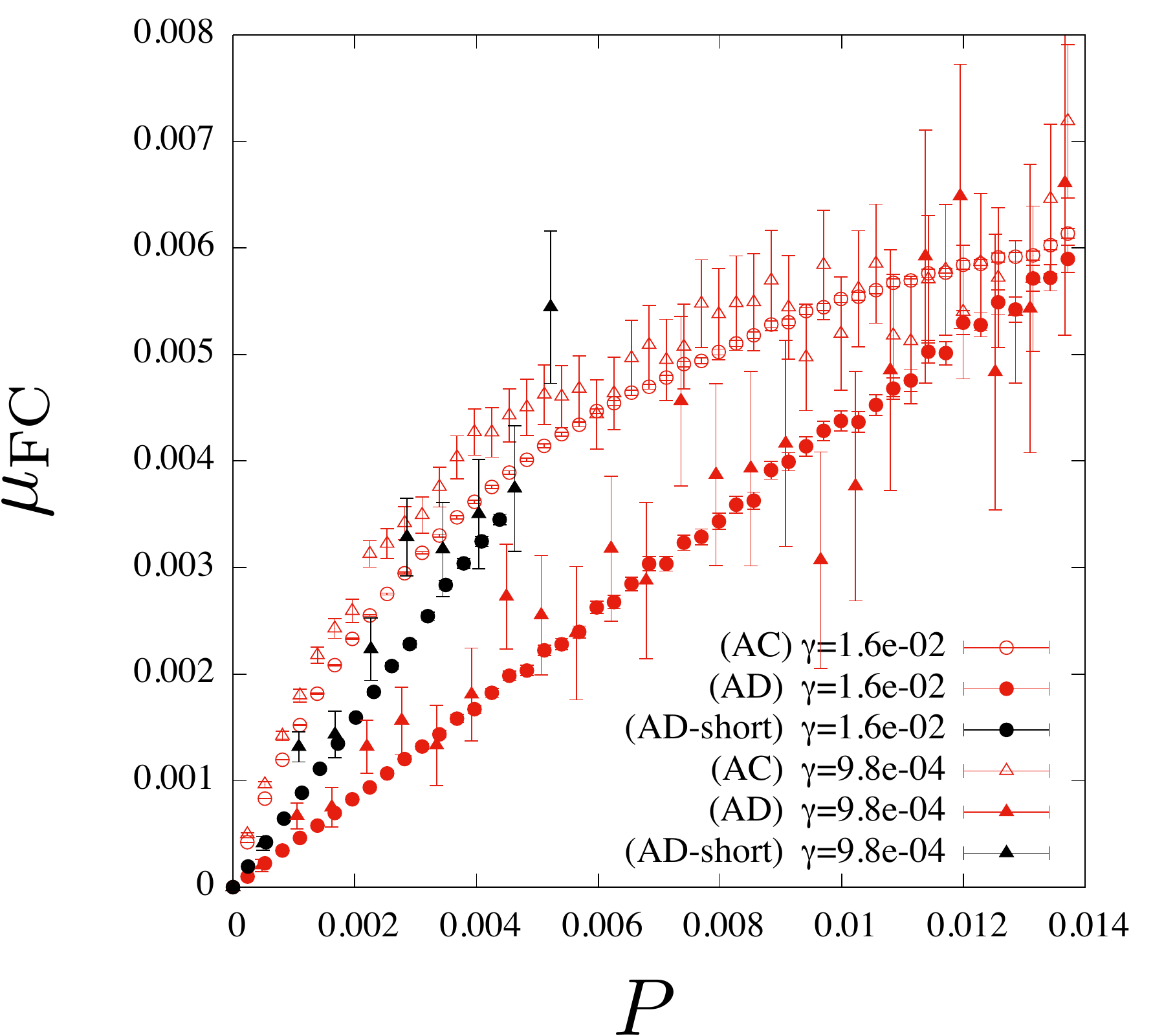}
  \caption{
  FC shear modulus measured under compression and decompression.
 Here data sets of $N=1000$ with  $\g=2^{-6} \simeq 1.6 \times 10^{-2}$ and $\g=2^{-8} \simeq 9.8 \times 10^{-4}$.
For the decompression, data obtained returning from $\varphi_{\rm f}=0.66$ ($P_{\rm f} \simeq 0.005$) (AD-short)
 and $\varphi_{\rm f}=0.68$ ($P_{\rm f} \simeq 0.014$) (AD) are shown.
 $dP=10^{-4}-5 \times 10^{-3}$ and the number of samples for each bin is $O(10^{5})$ for AC, $O(10^{4})$ for AD.
}
\label{fig:FC}
\end{figure}

\section{Results}

\subsection{Zero-field compression}

We first discuss results obtained with the ZFC protocol. We note that ZFC is the standard protocol that has been used in a number of previous
studies~\cite{OSLN03,IB15}, so we can directly compare our data with previous work.

In Fig.~\ref{fig:ZFC} we report results for $\mu_{\rm ZFC}$ obtained at constant pressure $P$ and for several values of shear strain $\g$.
We observe that at large $\g$ there is a strong non-linear contribution and $\mu_{\rm ZFC} \sim P$, but upon lowering $\g$ the linear
response regime emerges, because the curves converge towards the result obtained using the fluctuation formula. This result confirms that
$\mu_{\rm ZFC}$ is indeed a property of a single energy minimum, as it can be measured with the fluctuation formula while the system sits in
the minimum without applying any perturbation.

Also, $\mu_{\rm ZFC}$ is found, as in previous work~\cite{OSLN03}, to scale proportionally to $P^{1/2}$ when $P\to 0$. In order to have a clean demonstration
of this behavior, we collapse all curves at finite $\g$ using the 
form proposed in~\cite{GLS15}
in the framework of a proposed scaling theory of the jamming transition
:
\beq\label{eq:ZFCsc}
\mu_{\rm ZFC}(P,\g) = \g^{a/2} F(P/\g^a) \ .
\eeq
Here, $F(x\to\io) \sim x^{1/2}$, while $F(x\to 0) \sim x$. 
This implies $\mu_{\rm ZFC}(P,\g\to 0) \sim P^{1/2}$, while $\mu_{\rm ZFC}(P\to 0, \g>0) \sim P \g^{-a/2}$. In Fig.~\ref{fig:ZFCscaled}
we report a very good data collapse using the value of $a = 4/3$ proposed in~\cite{GLS15}. 
The very good coincidence with the prediction of~\cite{GLS15} confirms the validity of the scaling theory of jamming proposed there.
Furthermore,
this result confirms the scaling
of $\mu_{\rm ZFC}(P,\g=0)\sim P^{1/2}$~\cite{OSLN03} and extends it to the non-linear regime~\cite{GLS15,OH14,CSD14}.
Although finite size effects may appear at lower values of the pressure approaching the unjamming point, we checked that there is no signature of
finite size effects within the range of pressure we studied
down to $P=O(10^{-4})$ and system sizes $N=500-2000$.
Note {\it en passant} that, although this is not the main focus of this paper, the result in Eq.~\eqref{eq:ZFCsc}, proposed in~\cite{GLS15}, 
to the best of our knowledge has not
been tested numerically before, and therefore it is an original result of this work.

\subsection{Field compression}

We now turn to the discussion of the FC shear modulus, which
is reported in Fig.~\ref{fig:FC}.
First, we note that while for $\mu_{\rm ZFC}$ adiabatic compression (AC) and decompression (AD)
give identical results, this is not the case for $\mu_{\rm FC}$.
During AC, $\mu_{\rm FC}$ grows linearly at small $P$ and then saturates at larger $P$;
it remains larger during AC than during AD.
By comparing results for two different final values of pressure $P_{\rm f}$ (see Fig.~\ref{fig:FC}), we note
that the AD curves connects perfectly linearly the final value of stress just before
starting the decompression and $0$, 
which seems to be a general feature of the AD curves.
We also found that the AD curve is reversible, i.~e. the stress
follows exactly the AD curve under re-compression.

Moreover, while $\m_{\rm ZFC}$ deviates from the linear regime for quite small $\g$ ($\g\sim 10^{-4}$, see Fig.~\ref{fig:ZFC}), 
here we observe that $\m_{\rm FC}$ is almost independent of $\g$ in a regime of shear strain that is larger by two orders of magnitude,
 $\g \lesssim 10^{-2}$. 
We conclude that $\mu_{\rm FC}(P)$ is measured in the linear response regime, and is proportional to $P$ at low pressure
both in AC and AD protocols (although with different coefficients).

Having established the existence of a linear regime for both FC and ZFC shear moduli, in Fig.~\ref{fig:compare}
we compare the two. We find that at all pressures, $\mu_{\rm FC} < \mu_{\rm ZFC}$, with
$\mu_{\rm FC}\sim P \ll \mu_{\rm ZFC} \sim P^{1/2}$ in the jamming limit. 

\subsection{Comparison between glassy and crystalline solids}

\begin{figure*}[t]
\includegraphics[width=1.6\columnwidth]{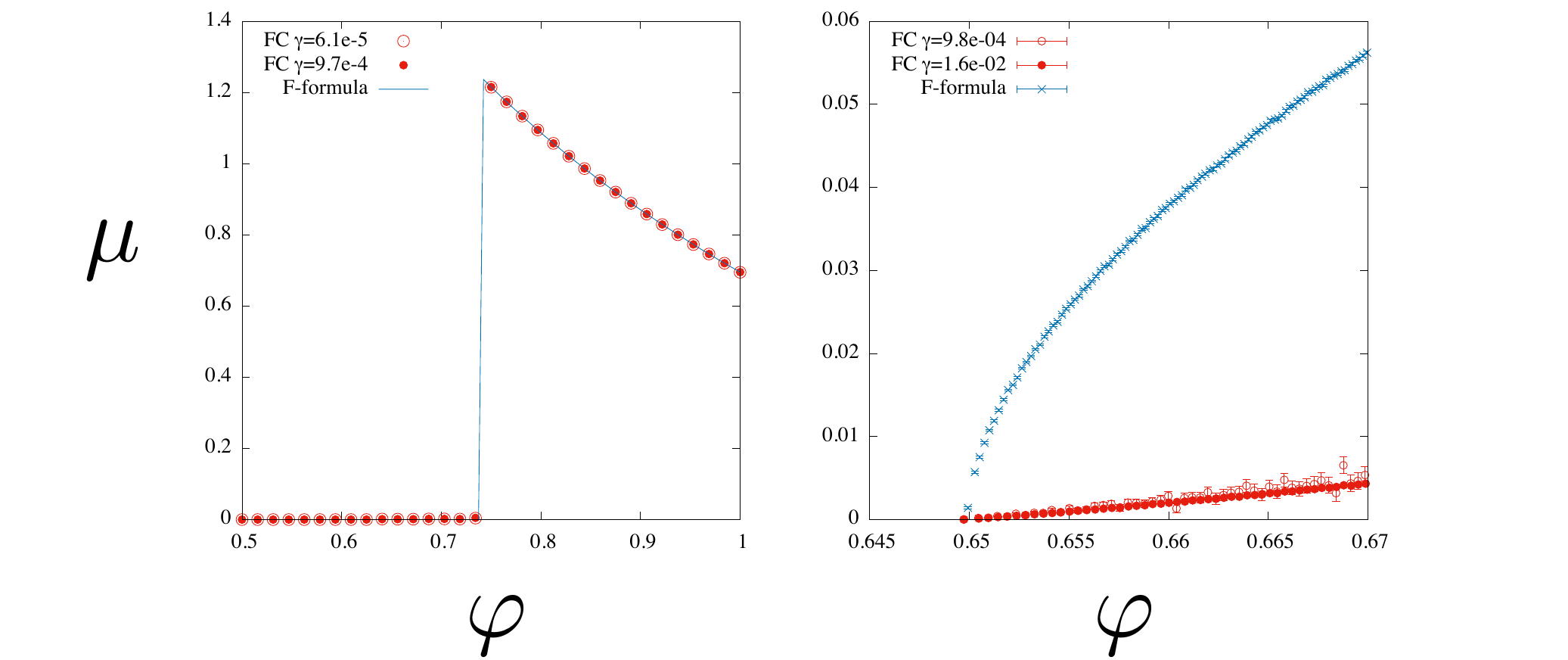}
\caption{
  Comparison of the ZFC (blue) and FC (red) shear modulus of the FCC crystal (left)
 and the glassy system (right).
 For the crystal and glassy systems $N=864$ and $N=1000$ are used respectively,
 and we use the AD protocol for the FC shear modulus of the glass.
 In both cases, the fluctuation formula is used to obtain the ZFC shear-modulus.
 For the glassy system, binning is done by $dP=5.0 \times 10^{-3}$
 for FC and $dP=10^{-4}$ for ZFC by which the number of samples for each bin becomes $O(10^{4})$.
$\mu_{\rm FC} < \mu_{\rm ZFC}$ for all pressures in the glassy system.}
\label{fig:compare}
\end{figure*}

In Fig.~\ref{fig:compare}, we 
compare the results for the amorphous solids with 
 the results for the FCC crystal, 
 whose close packing density is $\varphi_{\rm c}\simeq 0.72$. The ZFC shear modulus of the FCC system is obtained 
  using the fluctuation formula. We found that the non-affine correction
  term is absent in the FCC case.
On the other hand the FC shear modulus of the FCC system is obtained
performing precisely the same analysis as we did for the glassy system.
As expected, we find no difference between the FC and ZFC shear moduli in the crystalline system.

\subsection{Finite size effects}

\begin{figure}[t]
\includegraphics[width=0.8\columnwidth]{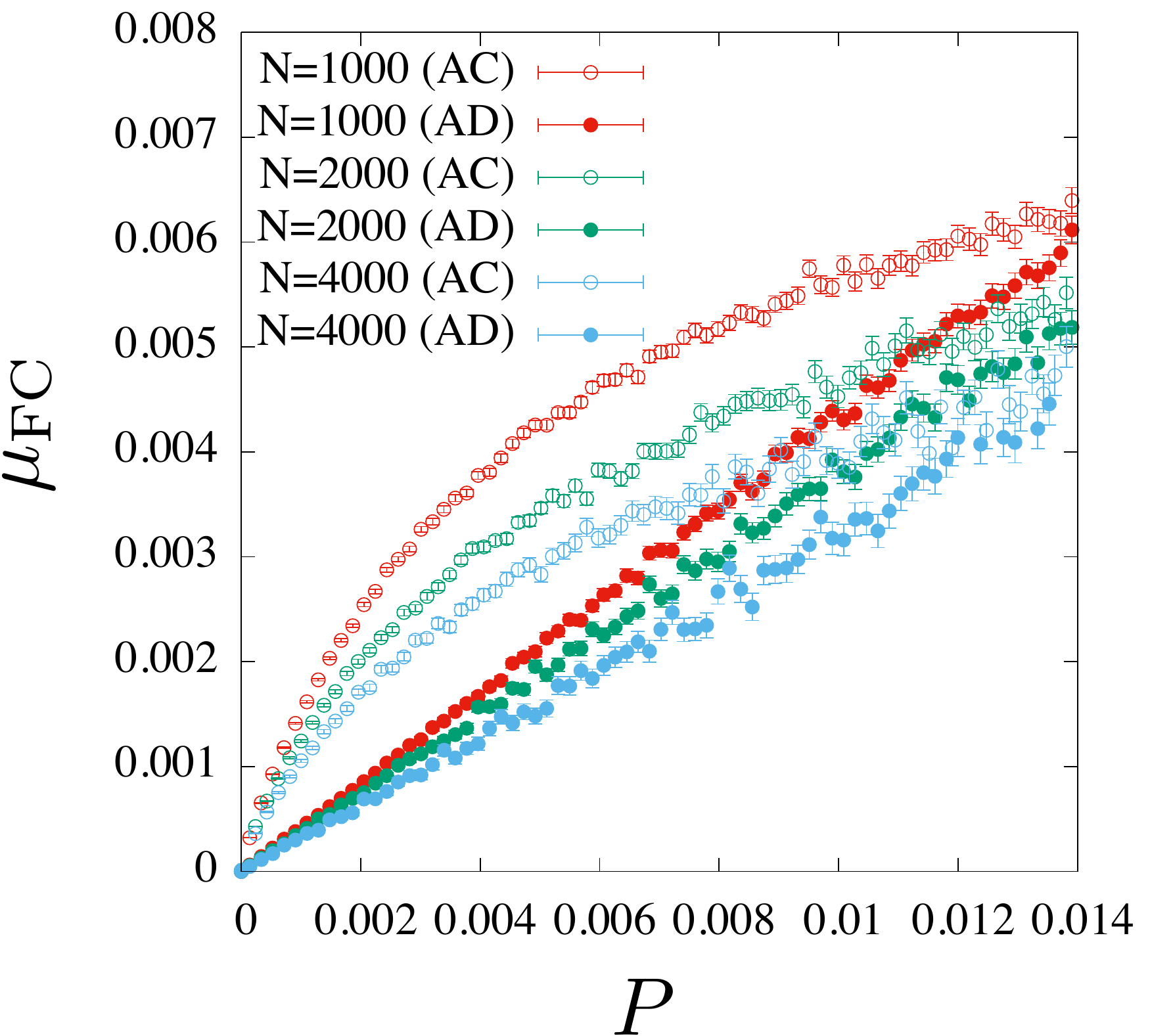}
\caption{
  Finite size effects of the FC shear modulus.
  $\g=2^{-6} \simeq 1.6 \times 10^{-2}$.
  $dP=10^{-4}$ and the number of samples for each bin is $O(10^{4})$.
}
\label{fig:FCfinitesize}
\end{figure}

\begin{figure}[t]
\includegraphics[width=0.8\columnwidth]{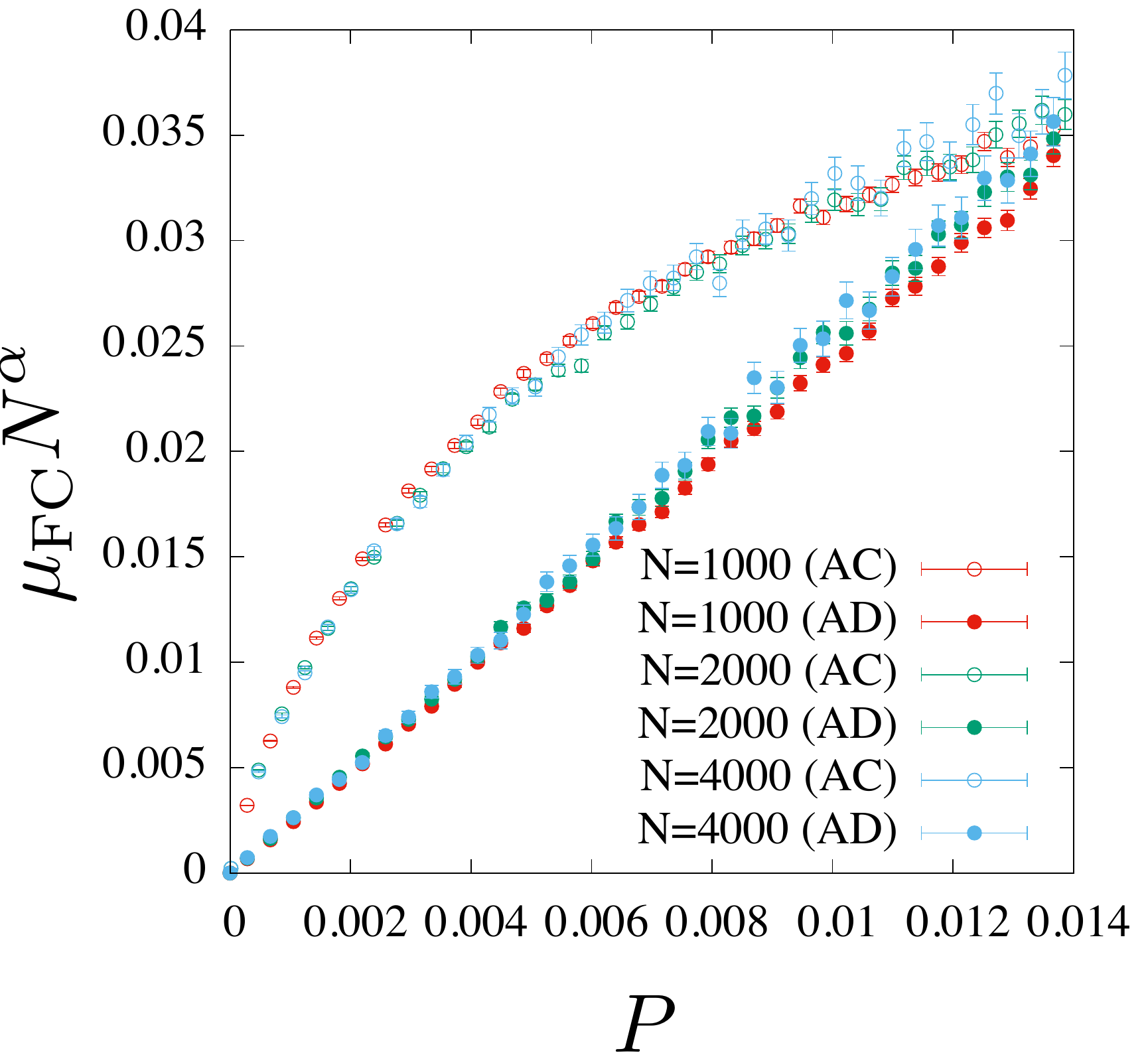}
\caption{
 Finite size scaling of the FC shear modulus.
 Here the data shown in  Fig.~\ref{fig:FCfinitesize} are used.
 The exponent is $\alpha=0.25$.
}
\label{fig:FCfinitesizescaling}
\end{figure}

In Fig.~\ref{fig:FCfinitesize} and Fig.~\ref{fig:FCfinitesizescaling} we study the system size dependence of $\m_{\rm FC}$.
We observe that $\m_{\rm FC}$ decreases upon increasing $N$ and that $N^{\a(\g)} \m_{\rm FC}$ is approximately
constant with a shear strain-dependent exponent $\a(\g)$, which suggests that $\m_{\rm FC}=0$ in the thermodynamic
limit for our samples.
However, we only obtained data on three sizes and the proposed $N^{-\alpha}$ behavior works 
with an exponent $\alpha$ that is not so big. 
There might be still the opportunity that the infinite size limit of $\mu_{\rm FC}$ 
is compatible also with a finite value, or that we are in a pre-asymptotic limit for the finite size scaling.
Finally, recall that we do not observe significant finite size effects for $\m_{\rm ZFC}$.

\section{Comparison with mean field theory}
\label{sec:theory}

As discussed in the introduction, the design of our numerical simulation 
has been inspired by the analytical solution of the infinite dimensional
problem, which provides a mean field theory for the problem and establishes an analogy with mean field spin glasses.
Although our work is not meant to be a precise test of the theory, it is still useful to compare our findings to what is 
expected on the basis of the mean field picture.

First of all, we have found that $\mu_{\rm FC} < \mu_{\rm ZFC}$, as suggested by the theory. Moreover,
upon approaching jamming, $\mu_{\rm FC} \sim P \ll \mu_{\rm ZFC} \sim P^{1/2}$, which is also consistent with
the theoretical expectation. The fact that $\mu_{\rm FC}$ is strongly dependent on the protocol (e.g. it is different for
AC and AD protocols, as shown in Fig.~\ref{fig:FC}) is also expected from the theory: because $\mu_{\rm FC}$ results
from an averaging over minima within a basin, its value depends on the way this minima are sampled out of equilibrium~\cite{YZ14}.
All these results are thus qualitatively consistent with the theoretical expectation.

There is one result, however, that deserves a more detailed discussion: the system size dependence of $\mu_{\rm FC}$.
Mean field theory predicts that glassy states prepared by slow annealing from the liquid should fall within
hierarchically-organised basins~\cite{RUYZ14,nature2}, having a FC shear modulus that remains finite
in the thermodynamic limit. Another theoretical approach is to put a uniform weight ({\it \`a la Edwards}) over all possible
glassy states: this leads to similar results~\cite{nature}.
Our preparation protocol instead starts from totally random {\it initial} configurations, which are instantaneously cooled
to very low temperature, and then to zero temperature (see Section~\ref{sec:preparation}): this is
very different from preparing
a glass by slow annealing or by giving the same weight to all possible glassy states~\cite{GBOS09,PZ10,TS10,BBCOS16}.
This is a possible reason that explains why $\mu_{\rm FC}$ is observed to vanish for $N\to\io$: 
our samples have been prepared very far from equilibrium
and therefore do not belong to fully stabilized glassy metabasins. How to compute properties of these states within mean
field theory remains an open question: it could be possible using dynamical methods such as the ones introduced in~\cite{MKZ15}.
It would thus be interesting to repeat this study using 
fully stabilised glasses for which theoretical predictions have been succesfully tested~\cite{nature2}: but this is much more
computationally demanding and we leave it for the future.
Another possibility that we cannot exclude is that specific finite-dimensional effects lead to the vanishing of $\mu_{\rm FC}$
in all glassy states: to test this idea, one should repeat
the present study in different spatial dimensions to see if a systematic trend with dimension emerges~\cite{CIPZ11}.

\section{Discussion}

In this paper we have shown that the shear modulus of a simple amorphous solid at zero temperature
is protocol dependent: there are at least two distinct
shear moduli in the linear response regime. 
The FC protocol, in which strain is applied before compression, 
leads to softer glasses than the ZFC protocol, in which strain is applied after compression.
The infinite-dimensional solution of the problem provides a natural interpretation of this result~\cite{YZ14,RU15}.
In the ZFC
protocol the system is first prepared in a minimum of the energy, then strain is applied. In this way
one probes the response of a single energy minimum. We confirm this by showing that $\mu_{\rm ZFC}$
can be equivalently obtained by the fluctuation formula, i.e. without applying strain but using linear response
in the vicinity of a single minimum.
In the FC instead, the strain is applied before compression, and during compression the system is allowed
to explore, through plastic events, some part of a larger ``basin'' composed by several energy minima. 
In this way more stress can be relaxed, leading to a softer response, 
$\mu_{\rm FC} < \mu_{\rm ZFC}$. 
Note that while
plastic events themselves 
are non-linear processes from the microscopic point of view,
they give rise 
to a ``renormalized'', softer linear response at the macroscopic level.
This result suggests the presence of at least two ``structures'' in the energy landscape: minima, and basins
of minima (Fig.~\ref{fig:land}). 

We also find that upon approaching the jamming point where pressure
vanishes, the ratio $\mu_{\rm FC} / \mu_{\rm ZFC} \propto P^{1/2}$ vanishes. This result is consistent with the
theoretical prediction obtained in infinite spatial dimensions where the structure of minima inside clusters is hierarchical
and fractal~\cite{nature,YZ14}. It thus hints at a very complex landscape characterized by many nested ``structures''.

Finally, we find that for the numerically investigated samples, $\m_{\rm FC} \to 0$ in the thermodynamic limit,
contrary to what the theory predicts for fully stabilized glassy basins. 
We tentatively attribute the discrepancy to the fact that theory focuses on fully
stabilized glasses, while in the numerical simulation we used glasses prepared from
totally random configurations, see the discussion in Section~\ref{sec:theory}.

Our results are related to other works and can be extended in several directions.
Explorations of plasticity in amorphous solids have been reported in many studies~\cite{ML99,CR00,ML04,RTV11,HJPS15}, 
where the instability
of energy minima under strain have been characterised in terms of soft energy modes~\cite{HKEP11,BW09b,Wy12,MW15}. 
In particular it has been suggested that plastic events happen for values of strain that vanish
when $N\to\io$ as power laws, $\d\g \sim N^{-\b}$~\cite{HKEP11,SBOS11,Wy12,MW15},
which suggests a non-trivial linear response
even in the vicinity of a single minimum.
It would be interesting to check whether this is consistent with our results and with theoretical predictions.
It is interesting to note that the cartoon in Fig.~\ref{fig:land} immediately suggests that
if one defines $\overline{\bullet}$ as the average over states, then
$d\overline{ \s}/d\g \neq \overline{ d\s/d\g }$, consistently with the results of~\cite{DPSS15}.
Furthermore, our results imply that there is dissipation even at zero frequency,
hence the dissipative part of the frequency-dependent 
shear modulus does not go to zero at low frequency, as in solid friction. 
This is one of the signatures of soft glassy rheology~\cite{SLHC97}, and is typical of energy landscapes with cusps 
like the one studied in~\cite{BBM96}.
Another interesting issue is that of non-linear responses, which are suggested to be strongly anomalous
both by theory~\cite{BU15} 
and numerical simulations~\cite{HKEP11,DPSS15,OH14}, 
in close relation with the complexity of the landscape suggested
by our results.
Finally, a crucial question is whether, upon adding temperature, the difference $\mu_{\rm FC} < \mu_{\rm ZFC}$ persists until the glass
melts, or there is a well defined temperature (a Gardner temperature) above which the glass becomes
a normal solid with $\mu_{\rm FC} = \mu_{\rm ZFC}$~\cite{nature,nature2}.

\acknowledgments

We warmly thank Jean-Philippe Bouchaud for bringing this problem to our attention and for many inspiring discussions,
and we thank two anonymous referees for suggesting important improvements
to a previous version of this manuscript.
We also thank Giulio Biroli, Hisao Hayakawa, Andrea Liu, Michio Otsuki, Giorgio Parisi, Corrado Rainone, Pierfrancesco Urbani
and Yuliang Jin for many useful discussions.
This work was supported by KAKENHI (No. 25103005  ``Fluctuation \& Structure'' and No. 50335337)
from MEXT, Japan, by JPS Core-to-Core Program ``Non-equilibrium dynamics of soft matter and informations''.
The Computations were performed using Research Center for Computational Science, Okazaki, Japan.

\end{document}